\documentclass[twocolumn,showpacs,preprintnumbers,amsmath,amssymb]{revtex4}

\usepackage{graphicx}
\usepackage{dcolumn}
\usepackage{bm}

\begin{document}

\title{Self Assembled II-VI Magnetic Quantum Dot as a Voltage-Controlled Spin-Filter.}

\author{C. Gould$^1$}
\author{A. Slobodskyy$^1$}%
\author{T. Slobodskyy$^1$}%
\author{P. Grabs$^1$}%
\author{D. Supp$^1$}%
\author{P. Hawrylak$^2$}%
\author{F. Qu$^2$}%
\author{G. Schmidt$^1$}%
\author{L.W. Molenkamp$^1$}%

\affiliation{%
$^1$Physikalisches Institut (EP3), Universit\"{a}t W\"{u}rzburg, Am
Hubland, D-97074 W\"{u}rzburg, Germany \\
$^2$Institute for Microstructural Sciences, NRC, Ottawa, K1A OR6,
Canada.
}%

\date{\today}

\begin{abstract}
A key element in the emergence of a full spintronics technology is
the development of voltage controlled spin filters to selectively
inject carriers of desired spin into semiconductors. We previously
demonstrated a prototype of such a device using a II-VI
dilute-magnetic semiconductor quantum well which, however, still
required an external magnetic field to generate the level splitting.
Recent theory suggests that spin selection may be achievable in
II-VI paramagnetic semiconductors without external magnetic field
through local carrier mediated ferromagnetic interactions. We
present the first experimental observation of such an effect using
non-magnetic CdSe self-assembled quantum dots in a paramagnetic
(Zn,Be,Mn)Se barrier.
\end{abstract}

\pacs{72.25.Dc, 85.75.Mm}
\maketitle

Nanomagnetics has over the past few years produced a series of
fascinating and often unanticipated phenomena. To name a few,
molecular magnets exhibit quantum tunnelling of the
magnetization\cite{1}, magnetic atoms on a surface exhibit giant
magnetic anisotropies\cite{2}, and magnetic domain walls are being
harnessed as data carriers\cite{3}. Here, we report on another
remarkable phenomenon: self-assembled quantum dots, fabricated from
II-VI dilute magnetic semiconductors (DMS) that macroscopically
exhibit paramagnetism, possess a remanent magnetization at zero
external field. This allows us to operate the dots as voltage
controlled spin filters, capable of spin-selective carrier injection
and detection in semiconductors. Such spin filter devices could
provide a key element in the emergence of a full spintronics
technology \cite{4}. We present the first experimental observation
of such a device using an approach based on the incorporation of
non-magnetic CdSe self assembled quantum dots (SADs) in paramagnetic
(Zn,Be,Mn)Se

We previously demonstrated a prototype of such a spin filter using a
II-VI DMS-based resonant tunnelling diode \cite{5}. However, while
that device was tuned by a bias voltage, the spin filtering
mechanism still required an external magnetic field. Moreover,
ferromagnetic III-V semiconductors like (Ga,Mn)As are not suitable
for resonant tunnelling devices due to the short mean free path of
holes \cite{6}. Recent theoretical works \cite{7,8,9} have suggested
that spin selection may be achievable in II-VI DMS without any
external magnetic field by creating localized carriers that might
mediate a local ferromagnetic interaction between nearby Mn atoms.

\begin{figure}
\centerline{\includegraphics[width=3.375in]{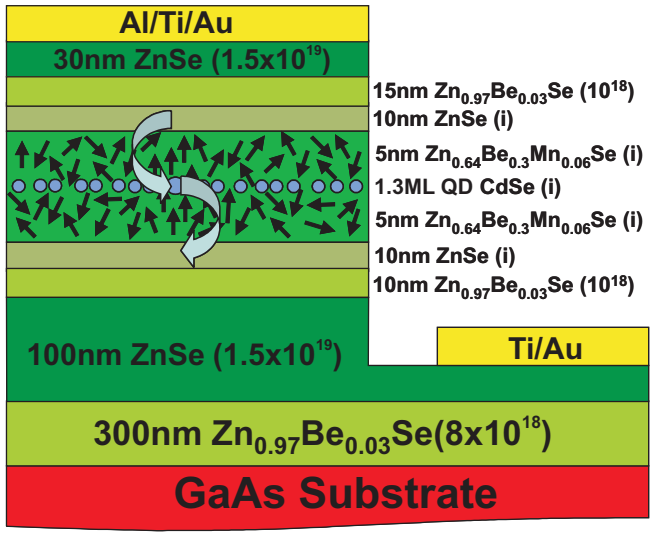}}
\caption{\label{mbe} Full layer structure of the device, and
schematic of the transport mechanism. As electrons tunnel through
the quantum dot, they mediate a local magnetic interaction between
nearby Mn ions causing them to align ferromagnetically.}
\end{figure}

Our sample is an MBE-grown all-II-VI resonant tunnelling diode (RTD)
structure consisting of a single 9 nm thick semi-magnetic
Zn$_{0.64}$Be$_{0.3}$Mn$_{0.06}$Se tunnel barrier, sandwiched
between gradient doped Zn$_{0.97}$Be$_{0.03}$Se injector and
collector. Embedded within the barrier are 1.3 monolayers of CdSe.
The lattice mismatch between the CdSe and the
Zn$_{0.64}$Be$_{0.3}$Mn$_{0.06}$Se induces a strain in the CdSe
material, which is relaxed by the formation of isolated CdSe dots
\cite{10}. The full layer stack is given in Fig.~\ref{mbe}. Standard
optical lithography techniques were used to pattern the structure
into 100 $\mu $m square pillars, and contacts were applied to the
top and bottom ZnSe layers in order to perform transport
measurements vertically through the layer stack. More details of the
fabrication procedure are given in Ref. \cite{5}. From the size of
the pillars, and the typical density of the dots, one would expect
some million dots within our device. However, despite this number,
transport through similar III-V SAD-RTDs is usually dominated by
only a few dots that come into resonance at lower bias voltages
\cite{11,12,13}. We therefore interpret the low bias transport
through our sample as corresponding to electrons tunnelling from the
injector into a single quantum dot and out of the dot into the
collector as schematically depicted in Fig.~\ref{mbe}. Based on
calculations of energy levels of strained quantum dots, we find
several quantum dot levels populated by electrons at zero bias.
Hence the electrons tunnel through excited states of quantum dots
containing a finite number of electrons.

\begin{figure}
\centerline{\includegraphics[width=3.375in]{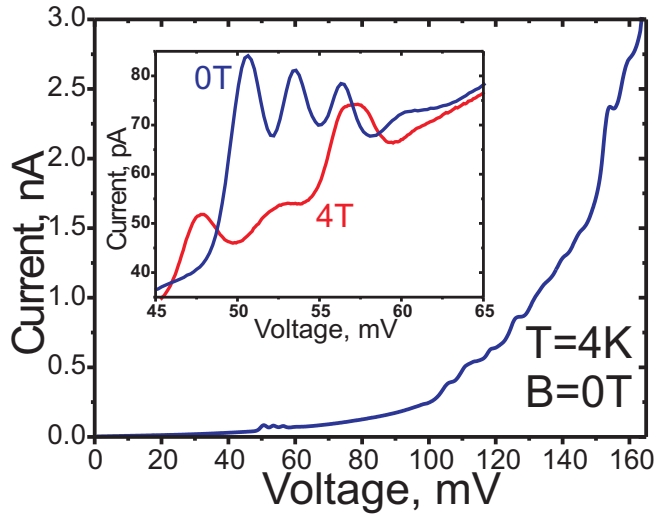}}
\caption{\label{IV} Current-voltage characteristic of the device,
with a high resolution view of the first resonance feature in the
inset, clearly showing a strong magnetic field dependence of the
resonances.}
\end{figure}

The experiments were carried out in a magnetocryostat and studied at
temperatures down to 1.3 K and in fields from 0 to 6 T.
Fig.~~\ref{IV} shows a full current voltage curve up to a bias
voltage of 170 mV. A first feature is observed at a bias of 55 mV,
associated with the first dot coming into resonance. At bias voltage
above 100 mV, several resonances due to the ensemble of dots can
also be observed. We will first focus on the low bias feature which
is shown in the inset to the figure. These more detailed curves
taken at 0 and 4T clearly show that the feature actually has a
complex structure, consisting of four distinct peaks, which evolve
with the magnitude of the applied magnetic field. We verified that
the evolution of the features does not appreciably depend on the
direction of the magnetic field, indicating that the magnetic
response of the system cannot be associated with artefacts such as
two dimensional states in the injector or wetting layer
\cite{5,11,12,13}, and that it must be a property of the dot or the
barrier. We also verified that the sample does not exhibit any
magnetic hysteresis.

\begin{figure}
\centerline{\includegraphics[width=3.375in]{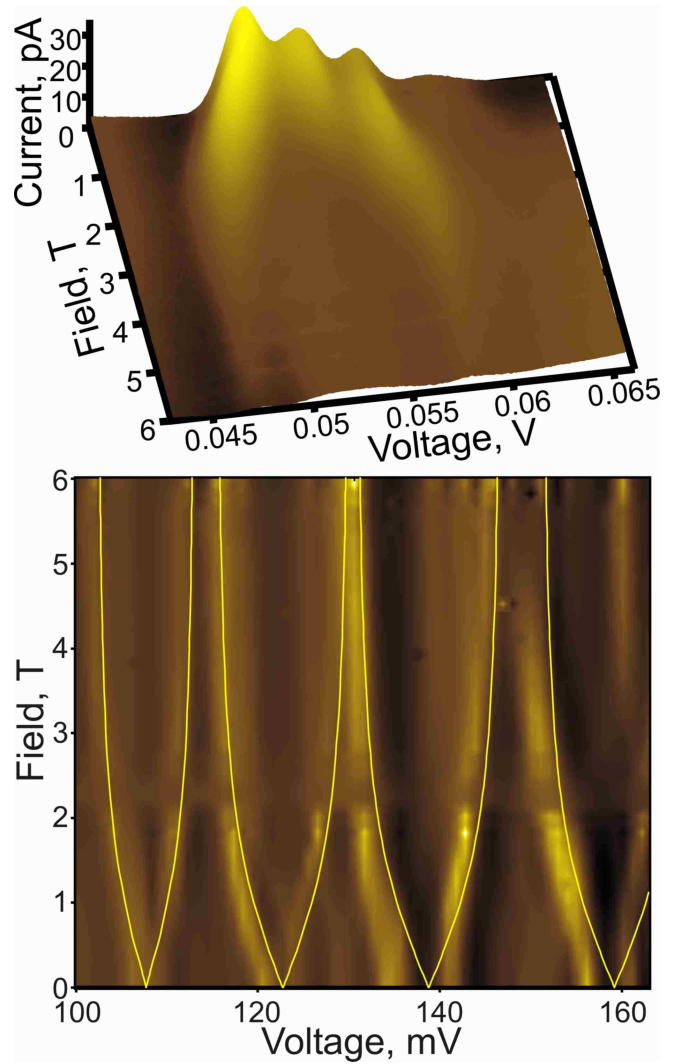}}
\caption{\label{Field_dependance} a) A surface plot of the current
through the device as a function of magnetic field and bias voltage
in the region of the first resonance feature. b) A colour scale
image of the resonances as a function of magnetic field and voltage
for higher bias resonances. Since these higher resonances are weaker
and on a significant background, the colour scale in b) is
proportional to the voltage derivative of the current in order to
better resolve the position of the resonances. In both a) and b),
the data at higher magnetic fields clearly has Brillouin like
behaviour, as evidenced in b), where Brillouin functions are plotted
as lines. However, at fields below ~500 mT, the behaviour departs
from a Brillouin function, with the splitting becoming constant and
remaining finite even at zero magnetic field.}
\end{figure}

A better understanding of the evolution of the features with
magnetic field can be obtained from Fig.~\ref{Field_dependance}. In
Fig.~\ref{Field_dependance}(a), we plot the current through the
device as a colour-scale surface with respect to bias voltage and
magnetic field. This puts into evidence two very important features
of the data. Firstly, that as the magnetic field is increased,
features split apart with a behaviour reminiscent of the Brillouin
function, and secondly, that the splitting remains finite in zero
external magnetic field. The same behaviour can be seen for many of
the higher bias resonances presented in
Fig.~~\ref{Field_dependance}(b). The first of these effects, that
the levels should split following a Brillouin function is not all
that surprising. It was previously observed in III-V devices that
resonances split in a magnetic field following the Land\'{e} g
factor of the material in the dots \cite{13}.

The main difference here is that in the present experiment, it is
the barrier, and not the dots that are magnetic. Since the effect of
the giant Zeeman splitting on the height of the barriers is
negligible, the presence of Mn should have little effect on the
barrier properties. However, given that electrons are not perfectly
localized in the dots, but rather have wave functions which extend
into the barrier, it is not surprising that the quantum levels in
the dots spin-split following the magnetization of the Mn in the
barriers, yielding results reminiscent of those previously observed
\cite{5} for the case of tunnelling through a dilute magnetic
quantum well.

The observation that the splitting remains finite at B=0 is however
more surprising since there is a priori nothing ferromagnetic in the
sample. This observation can be understood by considering the effect
of interactions between electrons in the dot and the Mn atoms in the
vicinity of the dot.

\begin{figure}
\centerline{\includegraphics[width=3.375in]{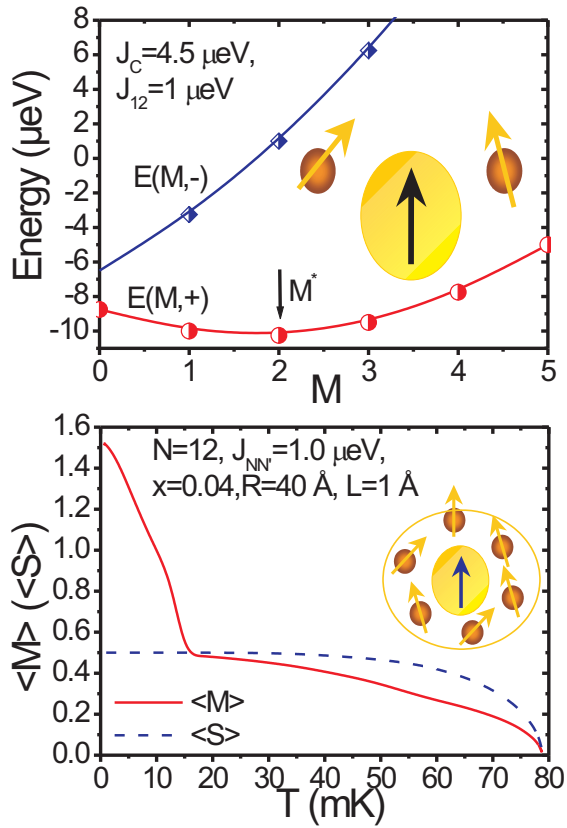}}
\caption{\label{Calculations} a) The energy levels E[+,M], E[-,M]
for two different electron spin orientation as a function of total
spin of two Mn ions M localized on a surface of spherical quantum
dot and b) average magnetization as a function of temperature  of Mn
ions randomly distributed on a surface of spherical quantum dot.}
\end{figure}

Electrons populate quantum dot levels according to the Pauli
exclusion principle, and Hunds rules \cite{14} whenever there is
orbital degeneracy. For a parabolic dot, the total electron spin
follows the sequence
S~=~{\{}1/2,0,1/2,1,1/2,0,1/2,1,3/2,{\ldots}{\}} with increasing
electron numbers. Hence, for almost all electron numbers, the total
spin of the dot is finite. The interaction of this total net spin
with the spin of Mn ions induces an effective ferromagnetic Mn-Mn
interaction. This can be seen by considering the total Hamiltonian
of the electronic and Mn system\cite{7,8,15,16}:
\begin{eqnarray*}
H=H_e +g^\ast \mu _B \vec {B}\cdot \sum\limits_i {\vec {S}_i } -J_C
\sum\limits_{\vec {R},i} {\vec {M}_{\bar {R}} \cdot } \vec {S}_i
\delta (\vec {r}_i -\vec {R})\\
+\sum\limits_{\vec {R}} {g_{Mn} \mu
_B \vec {B}\cdot \vec {M}_{\vec {R}} } +\frac{1}{2}\sum\limits_{\vec
{R},\vec {R}'} {J_{\vec {R},\vec {R}'} \vec {M}_{\vec {R}} \cdot }
\vec {M}_{\vec {R}'}
\end{eqnarray*}
Here $\vec {M}_{\vec {R}} $ is the spin of Mn ions ($M$=5/2) at
position $\vec {R}$, $S_{i}$ is the spin of the $i$-th electron
($S$=1/2). $J_{c}$ is the sp-d exchange constant between the
conduction electrons and the d-electrons of the Mn shell and
$J_{R{R}'} $ is the anti-ferromagnetic Mn-Mn interaction. The first
term is the spin independent Hamiltonian of electrons confined to a
quantum dot in a magnetic field, and interacting via a pair wise
potential. The full interaction between electron spins and Mn ions
in the barrier is an extremely complicated problem. We restrict
ourselves here to a demonstration that the electron spin is capable
of compensating the anti-ferromagnetic interaction among Mn ions and
lead to their ferromagnetic arrangement. We consider only a single
electron in the ground state and in the absence of external magnetic
field, leaving the problem of interacting many-electron dots for
future analysis. The effective spin Hamiltonian now reads:

\begin{equation}
\label{eq1} H=E_0 -J_C \sum\limits_{\vec {R}} {\vert \Phi (R)\vert
^2\vec {M}_{\bar {R}} \cdot } \vec {S}+\frac{1}{2}\sum\limits_{\vec
{R},\vec {R}'} {J_{\vec {R},\vec {R}'} \vec {M}_{\vec {R}} \cdot }
\vec {M}_{\vec {R}'}
\end{equation}

where $E_{0}$ is the electron energy and $\vert \Phi (R)\vert ^2$ is
the probability of finding an electron at the position $\vec {R}$ of
a Mn ion. Even for such a simplified Hamiltonian the number of
configurations is very large in the number of Mn ions. The physics
of Mn-Mn interactions mediated by electron spin can however be
understood by examining an exactly solvable problem of two
anti-ferromagnetically coupled Mn ions. The energy spectrum of the
coupled Mn-spin system is characterized by the total spin
J=M$\pm$1/2 where $M$ is the total Mn spin and the $\pm $1/2
corresponds to the directions of the electron spin. The evolution of
the energy of the system as a function of the total Mn spin depends
on the direction of the electron spin in the following way:
\begin{eqnarray*}
E(M,+)=-(\frac{\hat {J}_C }{2})M+(\frac{J_{N{N}'}
}{2})[M(M+1)-\frac{35}{2}] \\
E(M,-)=(\frac{\hat {J}_C }{2})(M+1)+(\frac{J_{N{N}'}
}{2})[M(M+1)-\frac{35}{2}]
\end{eqnarray*}
as shown in Fig.~\ref{Calculations}(a). In the absence of coupling
to the electron spin ($J_{c}$=0), it is obvious that the minimum
energy state for either electron spin corresponds to the total Mn
spin $M$=0, i.e. an antiferromagnetic arrangement. However, as shown
in Fig.~\ref{Calculations}(a), with coupling to the electron spin,
the $E(M,+)$ ground state of the combined system has finite total Mn
spin $M^{*} =(\frac{\hat {J}_C }{J_{N{N}'} }-1)/2$. To estimate the
value of $M^{*}$ we approximate our quantum dot by a spherical CdSe
dot with radius $R$= 4 nm and a barrier potential of 1 eV estimated
from strain and the Bir-Pikus Hamiltonian. The effective electron-Mn
exchange interaction for Mn ions on the surface of the sphere is
then given by $\hat {J}_C =J_C \vert \Phi (R)\vert ^2$ = 4.5 $\mu
$eV. For a typical Mn separation in the barrier of $R_{12}$=1.2 nm,
we estimate the antiferromagnetic interaction strength $J_{12}$= 1
$\mu $eV. Hence for our model system we find $M^{*}$= 2 and the
coupling to electron spin aligns spins of nearest neighbour Mn ions.
Independent mean field calculations involving tens of Mn ions
randomly distributed in the barrier around a spherical or disk
shaped quantum dots confirm the existence of ferromagnetic ordering
of Mn ions in the vicinity of quantum dots \cite{9}. In
Fig.~~\ref{Calculations}(b), we show the calculated averaged Mn and
electron spin magnetization as a function of temperature for Mn ions
localized in the barrier surrounding a spherical CdSe quantum dot
with radius of 4 nm and Mn concentration of 4{\%}. We find the
existence of the magnetic polaron, with the Mn magnetization
decaying as one moves away from the quantum dot. These finding are
in agreement with previous calculations of magnetic polarons
\cite{7,8,9}, and for reasonable parameters for our system shows
that the presence of electrons in the dot will mediate a local
ferromagnetic interaction between Mn atoms near this dot.

The interpretation of our experimental observations is therefore
clear. Electrons localized in the dot mediate a local ferromagnetic
interaction which causes a finite spin splitting even in the absence
of an external applied field. Our experiment is therefore tantamount
to measuring transport through a single magnetic polaron. The local
interaction has a strength corresponding to an effective field of
the order of some hundreds of mT, and can be randomly oriented. When
an external magnetic field is then applied, the ferromagnetic order
will first rotate towards the direction of the applied field, but
this will have no effect on the transport, which explains why in the
experimental data, the resonance positions are independent of
magnetic field for fields below $\sim $500 mT. However, as the
magnetic field is further increased, it will start to dominate, and
the spin splitting will grow following the normal paramagnetic
interaction of the dilute Mn system \cite{5}. A question remains as
to why the zero magnetic field splitting is observed here while it
was not seen in the optical measurements of Ref. \cite{16,17}. This
however can be understood by the fact that once current begins
flowing through the dot, a feedback mechanism sets in where spin
polarization of the current enhances the polarization of Mn spins
which in turn enhances the polarization of the current. \cite{7,8}
This dynamical effect also explains why spin polarization is
observed at much higher temperature than the predicted temperature
dependence of magneto-polaron of Fig.~~\ref{Calculations}(b).

In conclusion, we have shown that electrons in a quantum dot can
mediate a local ferromagnetic interaction in a surrounding dilute Mn
system, and that this leads to a finite energy splitting of spin
levels in the dot in the absence of an external magnetic field.
Coupled with the resonant tunnelling scheme which allows the bias
controlled selection of which dot level is used in tunnelling, our
results open up exciting new possibilities of a voltage controlled
spin filter which can operate in absence of any external magnetic
field, without relying on an inherent ferromagnetism of the
component materials.
\begin{acknowledgments}
The authors wish to thank V. Hock for help in sample fabrication,
and to acknowledge the financial support of ONR, DARPA, SPINOSA, and
the SFB
\end{acknowledgments}

\end{document}